\documentclass[3p]{elsarticle}
\usepackage{lineno,hyperref}
\usepackage{amsfonts}
\usepackage{float}
\usepackage{amsmath}
\usepackage{amssymb}
\usepackage{hyperref}
\usepackage{color}
\usepackage{graphicx}
\usepackage{amssymb}
\usepackage{amsmath}
\usepackage{graphicx}
\usepackage{dcolumn}
\usepackage{bm}
\usepackage{epsfig}
\usepackage[T1]{fontenc}
\usepackage{ae,aecompl}
\usepackage{booktabs}
\usepackage{tabularx}
\modulolinenumbers[10]

\journal{International Journal of Heat and Mass Transfer}

\begin{document}
\begin{frontmatter}

\title{Modeling incompressible thermal flows using a central-moment-based lattice Boltzmann method}

\author[mymainaddress]{Linlin Fei}

\author[mymainaddress,mysecondaryaddress]{K. H. Luo\corref{mycorrespondingauthor}}
\cortext[mycorrespondingauthor]{Corresponding author}
\ead{K.Luo@ucl.ac.uk}

\author[mymainaddress]{Chuandong Lin}
\author[mythirdaddress]{Qing Li}

\address[mymainaddress]{ Center for Combustion Energy; Key laboratory for Thermal Science and Power Engineering of Ministry of Education, Department of Thermal Engineering, Tsinghua University, Beijing 100084, China}

\address[mysecondaryaddress]{Department of Mechanical Engineering, University College London, Torrington Place, London WC1E 7JE, UK}

\address[mythirdaddress]{School of Energy Science and Engineering, Central South University, Changsha 410083, China}

\begin{abstract}
In this paper, a central-moment-based lattice Boltzmann (CLB) method for incompressible thermal flows is proposed. In the method, the incompressible Navier-Stokes equations and the convection-diffusion equation for the temperature field are sloved separately by two different CLB equations. Through the Chapman-Enskog analysis, the macroscopic governing equations for incompressible thermal flows can be reproduced. For the flow field, the tedious implementation for CLB method is simplified by using the shift matrix with a simplified central-moment set, and the consistent forcing scheme is adopted to incorporate forcing effects. Compared with several D2Q5 multiple-relaxation-time (MRT) lattice Boltzmann methods for the temperature equation, the proposed method is shown to be better Galilean invariant through measuring the thermal diffusivities on a moving reference frame. Thus a higher Mach number can be used for convection flows, which decreases the computational load significantly. Numerical simulations for several typical problems confirm the accuracy, efficiency, and stability of the present method. The grid convergence tests indicate that the proposed CLB method for incompressible thermal flows is of second-order accuracy in space.
\end{abstract}


\end{frontmatter}

\linenumbers

\section{Introduction}
As a mesoscopic numerical method based on the kinetic theory, the lattice  Boltzmann method (LBM) \cite{qian1992lattice,qian1995recent,chen1998lattice} has obtained remarkable success in the applications to fluid flows and heat transfer problems during the past three decades \cite{succi2001lattice,Gong2012A,Li2015Lattice,li2016lattice,Chen2016Double,Gong2017}. The LBM solves a  discreted Boltzmann equation, designed to recover the Navier-Stokes (N-S) equations in the macroscopic limit. The highly efficient and easy algorithm of LBM makes it at afforadable computatinal cost, while the meso-scale nature allows its natural
incorporation of micro and/or meso-scale physics \cite{li2016lattice}.

In the standard collision-streaming algorithm for LBM, the simplest collision operator is the single-relaxation-time (SRT) or BGK operator, in which all the distirbution functions are relaxed to their local equilibrium values at a common rate \cite{qian1992lattice}. However, the BGK-LBM may meet troubles of inaccuracy in implementing the boundary conditions \cite{ginzburg2003multireflection,pan2006evaluation,guo2008analysis}, as well as numerical instability at high Reynolds numebr or low-viscosity flows \cite{lallemand2000theory}. To overcome these difficulties, other collision operators, such as multiple-relaxation-time (MRT) operator \cite{d1994generalized,lallemand2000theory}, entropic operator \cite{karlin1999perfect,ansumali2000stabilization,ansumali2003minimal} and two-relaxation-time (TRT) operator \cite{ginzburg2005equilibrium,ginzburg2008two} have been developed.
Recently, a cascaded or central-moment-based operator was proposed by Geier et al. \cite{geier2006cascaded}. The collision in the central-moment-based Lattice Boltzmann method (CLBM) is carried out by relaxing central moments of the discrete distribution functions separately, rather than raw moments as in the MRT-LBM. By matching the higher order central moments of the continuous Maxwell-Boltzmann distribution naturally, CLBM achieves a higher  order of Galilean invariance \cite{geier2006cascaded}. Meanwhile, as mentioned by Geier et al. \cite{geier2006cascaded}, central moments can be expressed as polynomials of raw moments at the same order and below, which means relaxing raw moments (in MRT) affects the independent relaxation for central moments at higher orders. The ``cross-talk" is a source of numerical instability, and can be removed through relaxing central moments in CLBM. By setting high-oder central moments to their equilibrim values, CLBM has been used to simulate turbulence flow at $ Re=1400000 $ using coarse grids without resorting to any turbulence models or entropic stabilization \cite{geier2006cascaded}. More recently, CLBM has been extended to multiphase flows based on the interaction potential method \cite{shan1993lattice} by Lycett-Brown and Luo \cite{lycett2014multiphase} . Compared with the BGK-LBM for multiphase flows, the proposed multiphase CLBM enables significant improvment in reducing spurious currents near the phase interface \cite{lycett2014multiphase}, and achieving  higher stability range for the Reynolds number \cite{lycett2014binary}. They further extended the model with an improved forcing scheme \cite{lycett2015improved}, and made a breakthrough for large density ratio multiphase flow with high Reynolds and Weber numbers simultaneously \cite{lycett2016cascaded}.

Although CLBM has gained success in high Reynolds number single-phase flows and multiphase flows, its applications are still mainly limited to isothermal flows \cite{fei2016thermal}. The motivation of the present work is to extend CLBM to incompressible thermal flows. Up to now, the double-distribution function (DDF) approach has been extensively used for constructing thermal LBMs \cite{Chen2016Double,shan1997simulation,he1998novel,guo2002coupled,li2008improved,Wang2008Numerical,Gong2013Lattice,Yang2016Lattice}. In DDF-based incompressible thermal LBMs, a density or pressure distribution function is used to solve the velocity field, with another distribution for temperature field, where the two fields are usually coupled through the Boussinesq approximation \cite{guo2002coupled,li2008improved}. Due to the simplicity of the convection diffusion equation, a simpler lattice is often used for the temperature field  to decrease the computational load and save the memory storage \cite{guo2002coupled,mezrhab2010double}. The MRT operator is also widely used for the temperature field to achieve better numerical stability and more accuarte boundary conditions \cite{wang2013lattice,liu2016non,cui2016discrete}. Inspired by these studies, we try to extend the CLBM to incompressible thermal flows in the present study based on the DDF approach. Specifically, a density distribution function and a temperature distribution function are used to simulate the velocity field and the temperature field, respectively, and both of them are relaxed using the central-moment-based operater.

The rest of the paper is structured as follows: In Section \ref{sec.2},  the 2D CLBM for incompressible thermal flows is presented in detail.  Numerical experiments are carried out for several benchmark problems to validate the proposed method in  Section \ref{sec.3}. Finally, concluding remarks are given in Section \ref{sec.4}.

\section{CLBM for incompressible thermal flows}\label{sec.2}
In this section, details for the construction of the CLBM for incompressible thermal flows are given. The macroscopic govering equations for the flow fields are
\begin{subequations}\label{e1}
	\begin{equation}\label{e1a}
\nabla  \cdot {\bf{u}} = 0,
	\end{equation} 
	\begin{equation}
\frac{{\partial {\bf{u}}}}{{\partial t}} + {\bf{u}} \cdot \nabla {\bf{u}} =  - \frac{1}{{{\rho _0}}}\nabla p + \nu {\nabla ^2}{\bf{u}} + {\bf{F}},
	\end{equation} 
\end{subequations}
where $ {\bf{u}}  $, $ {p} $, $ {\rho _0} $, $ {\bf{F}} $ and $ \nu $ are the velocity, pressure, reference density, force field and kinematic viscosity, respectively. The convection-diffusion equation for a scalar variable $ \phi $ with diffusion coefficient $ D $ can be written as
\begin{equation}\label{e2}
\frac{{\partial \phi }}{{\partial t}} + {\bf{u}} \cdot \nabla \phi  = \nabla  \cdot (D\nabla \phi ).
\end{equation}
For the incompressible thermal flows considered in this study, the scalar variable and diffusion coefficient are specified as temperature $ T $ and thermal diffusivity $ \alpha 
$, respectively. To include the effect of temperature field on the flow field, the Boussinesq assumption is used and the force field is defined by
\begin{equation}\label{e3}
{\bf{F}} =  - {\bf{g}}\beta (T - {T_0}) + {{\bf{F}}_v},
\end{equation}
where the gravitational acceleration vector $ {\bf{g}} $ points to the  negative direction of y-axis, $ \beta $ is the thermal expansion coefficient, $ {T_0} $ is the reference temperature, and $ {\bf{F}}_v $ is an external body force.
\subsection{CLBM for the flow field}\label{sec.2.1}
The two-dimensional (2D) problems are considered in this study, and the D2Q9 lattice \cite{qian1992lattice} is used for the flow field. The lattice speed  $ c=\Delta{x}/\Delta {t}=1 $  is adopted, in which $ \Delta{x} $ and $ \Delta {t} $ are the lattice spacing and time step. The discrete velocities ${{\bf{e}}_i} = \left[ {\left| {{e_{ix}}} \right\rangle ,\left| {{e_{iy}}} \right\rangle } \right]
$ are defined by
\begin{subequations}\label{e4}
	\begin{equation}
	\left| {{e_{ix}}} \right\rangle  = {[0,1,0, - 1,0,1, - 1, - 1,1]^\top}, \\ 
	\end{equation} 
	\begin{equation}
	\left| {{e_{iy}}} \right\rangle  = {[0,0,1,0, - 1,1,1, - 1, - 1]^\top}, \\ 
	\end{equation} 
\end{subequations}
where $ i = 0...8 $, ${\left|  \cdot  \right\rangle }$ indicates the colunm vector, and the superscript $ \top $ indicates the transposition.

To construct the central-moment-based collision operator, raw moments and central moments for the discreted distribution functions 
(DFs) ${{f_i}}$ are introduced,
\begin{subequations}\label{e5}
	\begin{equation}
	{k_{{mn}}} = \left\langle {{f_i}\left| {e_{ix}^me_{iy}^n} \right.} \right\rangle , \\  
	\end{equation} 
	\begin{equation}
	{{\tilde k}_{{mn}}} = \left\langle {{f_i}\left| {{{({e_{ix}} - {u_x})}^m}{{({e_{iy}} - {u_y})}^n}} \right.} \right\rangle , \\
	\end{equation} 
\end{subequations}
and the equilibrium values $ k_{_{{mn}}}^{eq} $ and $ \tilde k_{{mn}}^{eq}$ are defined
analogously by replacing ${{f_i}}$ with the discrete equilibrium distribution functions (EDFs) $ {f_i^{eq}} $. In the literature, many researchers \cite{geier2006cascaded,lycett2014multiphase,lycett2014binary,lycett2016cascaded,fei2016thermal,fei2017consistent,premnath2009incorporating} using the recombined raw moments $ {{k_{20}} + {k_{02}},{k_{20}} - {k_{02}}} $, to treat the trace of the pressure tensor and the normal stress difference independently. To simplify the tedious implementation of using the recombined raw-moment set, a simplified method was proposed in \cite{Asinari2008Generalized}, where the raw moments ${{k_{20}},{k_{02}}}$ were used and some modifications were made in the relaxation matrix. In this work, the simplified raw-moment set is adopted,
\begin{equation}\label{e6}
\left| {{\Gamma_i}} \right\rangle  = 
\left[ {{k_{00}},{k_{10}},{k_{01}},{{k_{20}},{k_{02}}},{k_{11}},{k_{21}},{k_{12}},{k_{22}}} \right]^\top,
\end{equation}
and so do the recombined central moments $ \tilde \Gamma_i$. To be more specific, the raw moments can be given from ${{f_i}}$  through a transformation matrix $ {\bf{M}}$ by $ \left| {{\Gamma _i}} \right\rangle  = {\bf{M}}\left| {{f_i}} \right\rangle $, and the central moments shifted from raw moments can be realized through a shfit matrix $ {\bf{N}} $ by $ \left| {{{\tilde \Gamma }_i}} \right\rangle  = {\bf{N}}\left| {{\Gamma _i}} \right\rangle $. Similar to the expressions in \cite{fei2017consistent}, $ {\bf{M}}$ and $ {\bf{N}} $ are written as,
\begin{subequations}\label{e7}
	\begin{equation}
	{\bf{M}} = \left[ 
	\begin{array}{c c c c c c c c c}
	1 &1 &1&1&1&1&1&1&1\\
	0&1&0&-1&0&1&-1&-1&1\\
	0&0&1&0&-1&1&1&-1&-1\\
	0&1&0&1&0&1&1&1&1\\
	0&0&1&0&1&1&1&1&1\\
	0&0&0&0&0&1&-1&1&-1\\
	0&0&0&0&0&1&1&-1&-1\\
	0&0&0&0&0&1&-1&-1&1\\
	0&0&0&0&0&1&1&1&1\\
	\end{array} 
	\right], 
	\end{equation} 
	\begin{equation}
	{\bf{N}} = \left[ 
	\begin{array}{c c c c c c c c c}
	1 &0 &0&0&0&0&0&0&0\\
	-{u_x}&1&0&0&0&0&0&0&0\\
	-{u_y}&0&1&0&0&0&0&0&0\\
	u_x^2&-2{u_x}&0&1&0&0&0&0&0\\
	u_y^2&0&2{u_y}&0&1&0&0&0&0\\
	{u_x}{u_y}&-u_y&-u_x&0&0&1&0&0&0\\
	-u_x^2{u_y}&2{u_x}{u_y}&u_x^2&- {u_y}	
	&0&-2u_x&1&0&0\\
	-u_y^2{u_x}&{u_y}^2&2{u_x}{u_y}&0	
	&-{u_x}&-2u_y&0&1&0\\
	u_x^2u_y^2&-2{u_x}u_y^2&-2{u_y}u_x^2&u_y^2&u_x^2&4{u_x}{u_y}&-2{u_y}&-2{u_x}&1\\
	\end{array} 
	\right].
	\end{equation} 
\end{subequations}
By relaxing each central moment to its equilibrium counterpart independently, the post-collision central moments are given by
\begin{equation}\label{e8}
\begin{aligned}
\left| {\tilde \Gamma_i^*} \right\rangle = ({\bf{I - S}})\left| {{{\tilde \Gamma}_i}} \right\rangle  + {\bf{S}}\left| {\tilde \Gamma_i^{eq}} \right\rangle  + ({\bf{I - S}}/2)\left| {{C_i}} \right\rangle, \\   
\end{aligned}
\end{equation}
where $ {C_i} $ are the forcing source terms in central moments space, and the block-diagonal relation matrix is given by,
\begin{equation}
{\bf{S}} = diag\left( {[0,0,0],\left[ \begin{array}{l}
	{s_ + },{s_ - } \\ 
	{s_ - },{s_ + } \\ 
	\end{array} \right],[{s_v},{s_3},{s_3},{s_4}]} \right),
\end{equation}
with ${s_ + } = ({s_b} + {s_\nu})/2$ and ${s_ - } = ({s_b} - {s_\nu})/2$.
The equalibrium central moments of the $ {f_i^{eq}} $ are set equal to the continuous central moments of the Maxwellian-Boltzmann distribution in continuous velocity space,
\begin{equation}\label{e9}
\left| {\tilde \Gamma_i^{eq}} \right\rangle  = \left[ {\rho ,0,0,\rho c_s^2,\rho c_s^2,0,0,0,\rho c_s^4} \right]^\top,
\end{equation}
where $ \rho $ is the fluid density, and $ c_s=\sqrt{1/3} $ is the lattice sound speed. 
The corresponding EDF is in fact a generalized local equilibrium \cite{Asinari2008Generalized,premnath2009incorporating}. Consistently, the forcing source terms in central moments space are given by \cite{fei2017consistent},
\begin{equation}\label{e11}
\left| {{C_i}} \right\rangle  = {\bf{NM}}\left| {{R_i}} \right\rangle  = {[0,{F_x},{F_y},0,0,0,c_s^2{F_y},c_s^2{F_x},0]^ \top }.
\end{equation}

In the  streaming step, the post-collision discrete DFs in space $ \bf{x} $ and time $ t $ stream to their neighbors in the next time step as usual
\begin{equation}\label{e12}
{f_i}(\textbf{x} + {\textbf{e}_i}\Delta t,t + \Delta t) = f_i^*(\textbf{x},t),
\end{equation}
where the post-collision discrete DFs are determined by $ \left| {f_i^*} \right\rangle  = {{\bf{M}}^{ - 1}}{{\bf{N}}^{ - 1}}\left| {\tilde \Gamma_i^*} \right\rangle$.
Using the Chapman-Enskog analysis, the incompressible N-S equaltions in Eq. (\ref{e1}) can be reproduced in the low-Mach number limit \cite{premnath2009incorporating}. The hydrodynamics variables are obtained by,
\begin{equation}\label{e13}
\rho  = \sum\nolimits_i {{f_i}} ,~~~\rho {\bf{u}} = \sum\nolimits_i {{f_i}} {{\bf{e}}_i} + \frac{{\Delta t}}{2}{\bf{F}}.
\end{equation}
The kinematic and bulk viscosities are related to the relaxation parameters by $\nu  = (1/{s_\nu} - 0.5)c_s^2\Delta t $ and $ \xi  = (1/{s_b} - 0.5)c_s^2\Delta t $, respectively.
\subsection{CLBM for the temperature field}\label{sec.2.2}
A new D2Q5 (the five discrete velocity set is defined in Eq. (\ref{e4}), $ \left\{ {{{\bf{e}}_i} = \left[ {\left| {{e_{ix}}} \right\rangle ,\left| {{e_{iy}}} \right\rangle } \right]\left| {i = 0,1,...4} \right.} \right\}
 $) CLBM is proposed to solve the convection-diffusion equation for the temperature field. Similarly, the raw moments and central moments of the temperature distribution functions $ {g_i} $ can be defined by 
\begin{subequations}\label{e14}
	\begin{equation}
	{k_{mn}^T = \left\langle {{g_i}\left| {e_{ix}^me_{iy}^n} \right.} \right\rangle ,} 
	\end{equation} 
	\begin{equation}
	 {\tilde k_{mn}^T = \left\langle {{g_i}\left| {{{({e_{ix}} - {u_x})}^m}{{({e_{iy}} - {u_y})}^n}} \right.} \right\rangle}.
	\end{equation} 
\end{subequations}
In the D2Q5 lattice, the following five raw moments are considered,
\begin{equation}\label{e15}
\left| {\Gamma _i^T} \right\rangle  = {\left[ {k_{00}^T,k_{10}^T,k_{01}^T,k_{20}^T, k_{02}^T} \right]^ \top },
\end{equation}
and so do the recombined central moments $ \left| {\tilde \Gamma _i^T} \right\rangle $.
Analogously, the raw moments and central moments can be calculated through a transformation matrix $ {{{\bf{M}}_\textbf{T}}} $ and a shift matrix $ {{{\bf{N}}_\textbf{T}}}$,respectively
\begin{equation}\label{e16}
{\left| {\Gamma _i^T} \right\rangle  = {{\bf{M}}_\textbf{T}}\left| {{g_i}} \right\rangle,}~~~
{\left| {\tilde \Gamma _i^T} \right\rangle  = {{\bf{N}}_\textbf{T}}\left| {\Gamma _i^T} \right\rangle .}
\end{equation} 
Explicitly, the transformation matrix $ {{{\bf{M}}_\textbf{T}}} $ is expressed as
\begin{equation}\label{e17}
{{\bf{M}}_\textbf{T}}= \left[ 
\begin{array}{c c c c c}
1 &1 &1&1&1\\
0&1&0&-1&0\\
0&0&1&0&-1\\
0&1&0&1&0\\
0&0&1&0&1\\
\end{array} 
\right],
\end{equation}
and the shift matrix $ {{\bf{N}}_\textbf{T}} $ is given by
\begin{equation}\label{e18}
{{\bf{N}}_\textbf{T}} = \left[ 
\begin{array}{c c c c c}
1 &0 &0&0&0\\
-{u_x}&1&0&0&0\\
-{u_y}&0&1&0&0\\
u_x^2&-2{u_x}&0&1&0\\
u_y^2&0&-2{u_y}&0&1\\
\end{array} 
\right].
\end{equation} 
The collision in central moments can also be written as
\begin{equation}\label{e19}
\left| {\tilde \Gamma _i^{T,*}} \right\rangle  = ({\bf{I}} - {{\bf{S}}_\textbf{T}})\left| {\tilde \Gamma _i^T} \right\rangle  + {{\bf{S}}_\textbf{T}}\left| {\tilde \Gamma _i^{T,eq}} \right\rangle, 
\end{equation} 
where ${{\bf{S}}^\textbf{T}} = diag({\lambda _o},{\lambda _1},{\lambda _1},{\lambda _2},{\lambda _2})$ is the diagonal relaxation matrix. The equibrium values of the central moments are given by
\begin{equation}\label{e20}
\left| {\tilde \Gamma _i^{T,eq}} \right\rangle  = {\left[ {T,0,0,Tc_{T,s}^2,Tc_{T,s}^2} \right]^ \top },
\end{equation} 
where $ c_{T,s}^2 $ is the sound speed in the D2Q5 lattice. The post-collision temperature distribution functions $ g_i^*$ can be obtained by
\begin{equation}\label{e21}
g_i^* = {\bf{M}}_\textbf{T}^{ - 1}{\bf{N}}_\textbf{T}^{ - 1}\left| {\tilde \Gamma _i^{T,*}} \right\rangle. 
\end{equation} 
The streaming step for $ g_i^*$ are also as usual
\begin{equation}\label{e22}
{g_i}({\bf{x}} + {{\bf{e}}_i}\Delta t,t + \Delta t) = g_i^*({\bf{x}},t). 
\end{equation}
The temperature $ T $ is computed as
\begin{equation}\label{e23}
T = \sum\nolimits_{i = 0}^4 {{g_i}}.
\end{equation}
As shown in Appendix A, the convection-diffusion Eq. (\ref{e2}) can be recovered by the D2Q5 CLBM presented in this subsection, and the thermal diffusivity is related to the relaxation parameter by $\alpha  = (1/{\lambda _1} - 0.5)c_{T,s}^2\Delta t$.
\section{Numerical experiments}\label{sec.3}
In this section, we conduct several benchmark cases to verify the effectiveness and accuracy of the proposed CLBM for incompressible thermal flows. Unless otherwise specified, the lattice sound speed for the D2Q5 CLBM is set to $ {c_{sT}} = 1/2$, the tunable relaxation parameters for high-order central moments are set to 1.0, and the the non-equilibrium bounce-back method \cite{zou1997pressure} and non-equilibrium extrapolation method \cite{guo2007thermal} are adopted for velocity and temperature boundary conditions in simulations, respectively.
\subsection{The decay of a temperature wave}\label{sec.3.1}
\begin{figure*}[!ht]
	\center {
		{\epsfig{file=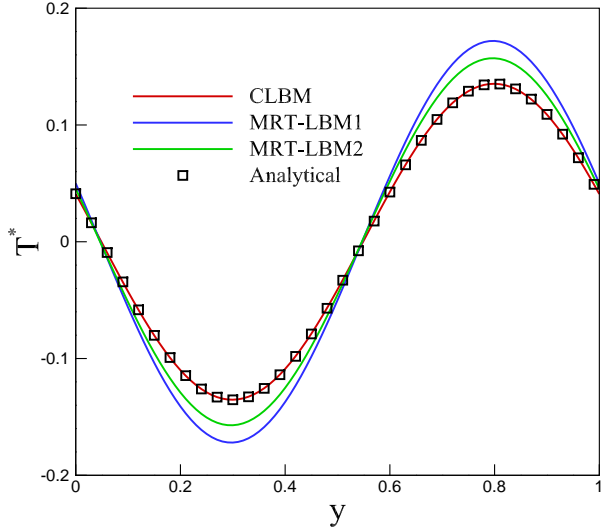,width=0.5\textwidth,clip=}}	
	}
	\caption{Comparison of the temperature profiles at $ {t^*} = 2.0 
		$ simulated by different methods.}
	\label{FIG1a}
\end{figure*}

\begin{figure*}[!ht]
	\center {
		{\epsfig{file=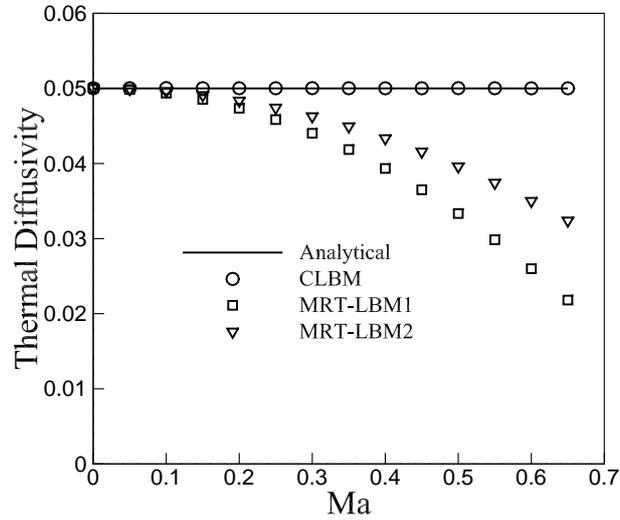,width=0.5\textwidth,clip=}}	
	}
	\caption{Comparison of the measured thermal diffusivities of each method at different values of $ Ma $.}
	\label{FIG1b}
\end{figure*}

Firstly, the decay of a temperature wave on a moving frame is considered. The problem is specified by the following velocity and temperature fields:
\begin{subequations}\label{eq24}
	\begin{equation}
	{\bf{u}} = [0,A].
	\end{equation} 
	\begin{equation}
	T = {T_0} + B\sin \left[ {\phi (y - At)} \right]\exp ( - {\phi ^2}\alpha t).	
	\end{equation} 
\end{subequations}
where $ \phi  = 2\pi /L$, A represents the vertical reference velocity component, and $ B $ is the initial amplitude of the temperature wave. Periodic boundary conditions are used along the $ x $ and $ y $ axes. In the simulations, $ {T_0} = 1.0 $, $ L = 1{\rm{00}} $, and the initial amplitude is set to $ B = 0.01 $. The velocity field is given, thus only the D2Q5 CLBM is adopted to solve the temperature field. Another two D2Q5 MRT LBMs in \cite{liu2016non} and \cite{cui2016discrete} are also used for comparison, and they are denoted by MRT-LBM1 and MRT-LBM2, respectively. Firstly, the case at Mach number $ {\rm{Ma = }}{A}/{c_s}=0.3 $ is considered. The profiles for the dimensionless temperature $ {T^*} $ in different methods at the time $ {t^*} = 2.0 $ are shown in Fig. \ref{FIG1a}, where $ {T^*} = (T - {T_0})/B $ and $ {t^*} = {\phi ^2}\alpha t$. It is found that the simulation result of the present CLBM is in good agreement of the analytical solution, while there are visible differences between the analytical solution and numerical solutions by the other two methods.
The measured thermal diffusivity of the simulated fluid is obtained by measuring the time decay of the temperature wave. Then, the measured thermal diffusivities of each method at different values of the Mach number are compared in Fig. \ref{FIG1b}, while the originally given thermal diffusivity is $ \alpha = 0.05 $.
For the present D2Q5 CLBM, the measured thermal diffusivity is independent of 
the reference velocity (or $ Ma $) and always agrees with the given value. For the other two D2Q5 MRT LBMs, the measured diffusivities decrease with the increase of the reference velocity. To be specific, the relative errors at $ Ma=0.3 $ are around $ 12\% $ and $ 8\% $ for MRT-LBM1 and MRT-LBM2, respectively.
\subsection{Normal plate velocity problem with a temperature difference}\label{sec.3.2}
The normal plate velocity problem is a fully developed channel flow, where the upper plate moves with a uniform velocity $ {u_0} $, and a uniform normal flow with velocity $ {v_0} $ is injected through the bottom plate  and withdrawn from the upper plate. The analytical solution of the flow is given by \cite{noble1995consistent},
\begin{equation}
\frac{{{u_a}}}{{{u_0}}} = \frac{{\exp (Re \cdot y/L){\rm{ - }}1}}{{\exp (Re){\rm{ - }}1}},
\end{equation} 
where the Reynolds number is based on the width of the channel, $ L $, and defined by $ {\mathop{ Re}\nolimits}  = {v_0}L/\nu $. For the present study, a temperature difference
$\Delta T = {T_H} - {T_L} $ is considered, where $ {T_H} $ and $ T_L $ are the temperatures at upper hot plate and bottom cold plate, respectively. The steady temperature profile satisfies \cite{guo2002coupled},
\begin{equation}
\frac{{{T_a} - {T_L}}}{{\Delta T}} = \frac{{\exp (RePr \cdot y/L){\rm{ - }}1}}{{\exp (RePr){\rm{ - }}1}},
\end{equation} 
where $Pr  = \nu /\alpha $ is the Prandtl number.

In the simulations, $ T_H=1.0 $ and $ T_L=0 $ are used, and the viscosity is set to $ \nu=0.1 $. Firstly, we set the Reynolds number $ Re=10 $, $ {u_0} = {v_0} = {\mathop{Re}\nolimits} \nu /L$ with $ L = 50 $.
Periodic boundary conditions are adopted at the inlet and outlet of channel, and the length of the channel is covered by 5 grids to save the computational cost. Three simulation cases with $Pr  = [0.1,1,10] $ are conducted to verify the numerical performace of the present method at a wide range of the Prandtl number. The tunable parameter $ {\lambda _2} $ is set according to the zero-numerical-slip condition \cite{cui2016discrete}, ${\lambda _2} = 12({\lambda _1} - 2)/({\lambda _1} - 12)$.
\begin{figure*}[!ht]
	\center {
		{\epsfig{file=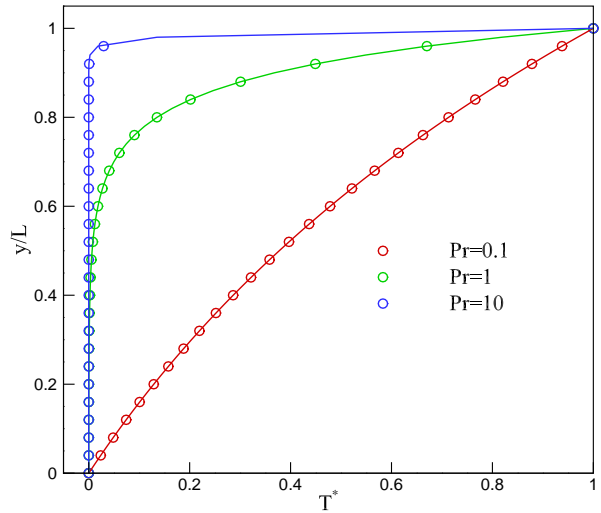,width=0.5\textwidth,clip=}}	
	}
	\caption{Comparison of numerical temperature profiles (symbols) and the analytical solutions (solid lines) at different  values of the $ Pr $.}
	\label{FIG2a}
\end{figure*}
The residual error $ {E_R} < {1\times10^{ - 9}} $ is used as the convergent criterion, the defination of $ {E_R} $ can be seen in \cite{fei2017consistent}. As shown in Fig. \ref{FIG2a},
the numerical results for the non-dimensional temperature $ {T^*} = (T - {T_0})/\Delta T $ agree very well with the analytical solutions. 

Then we set the Prandtl number corresponding to air, $Pr  =0.71 $, with different values of the Reynolds number, $ Re  = [10,20,30] $. The width of the channel covered by a series of grid nodes, $L  = [30,60,90,120,150] $, are considered to validate the convergence rate in space. The relative errors of temperature and velocity are calculated according to the following definations,
\begin{equation}
{E_T} = \sqrt {\frac{{\sum {{(T - {T_a})}^2}}}{{\sum {T_a}^2}}}, ~~{E_u} = \sqrt {\frac{{\sum {{(u - {u_a})}^2}}}{{\sum {u_a}^2}}}.
\end{equation}
\begin{figure*}[!ht]
	\center {
		{\epsfig{file=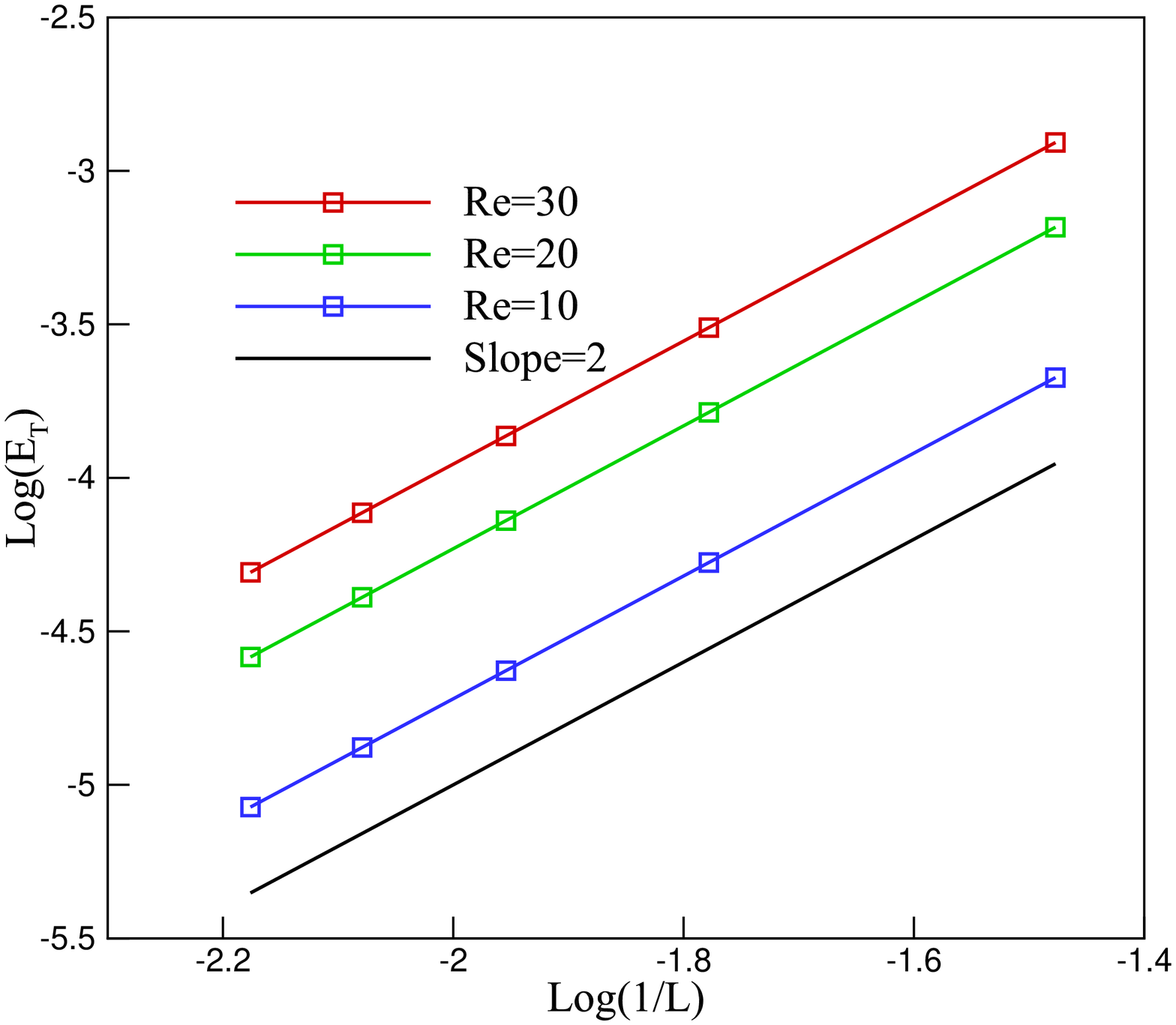,width=0.4\textwidth,clip=}}\hspace{0.5cm}  
		{\epsfig{file=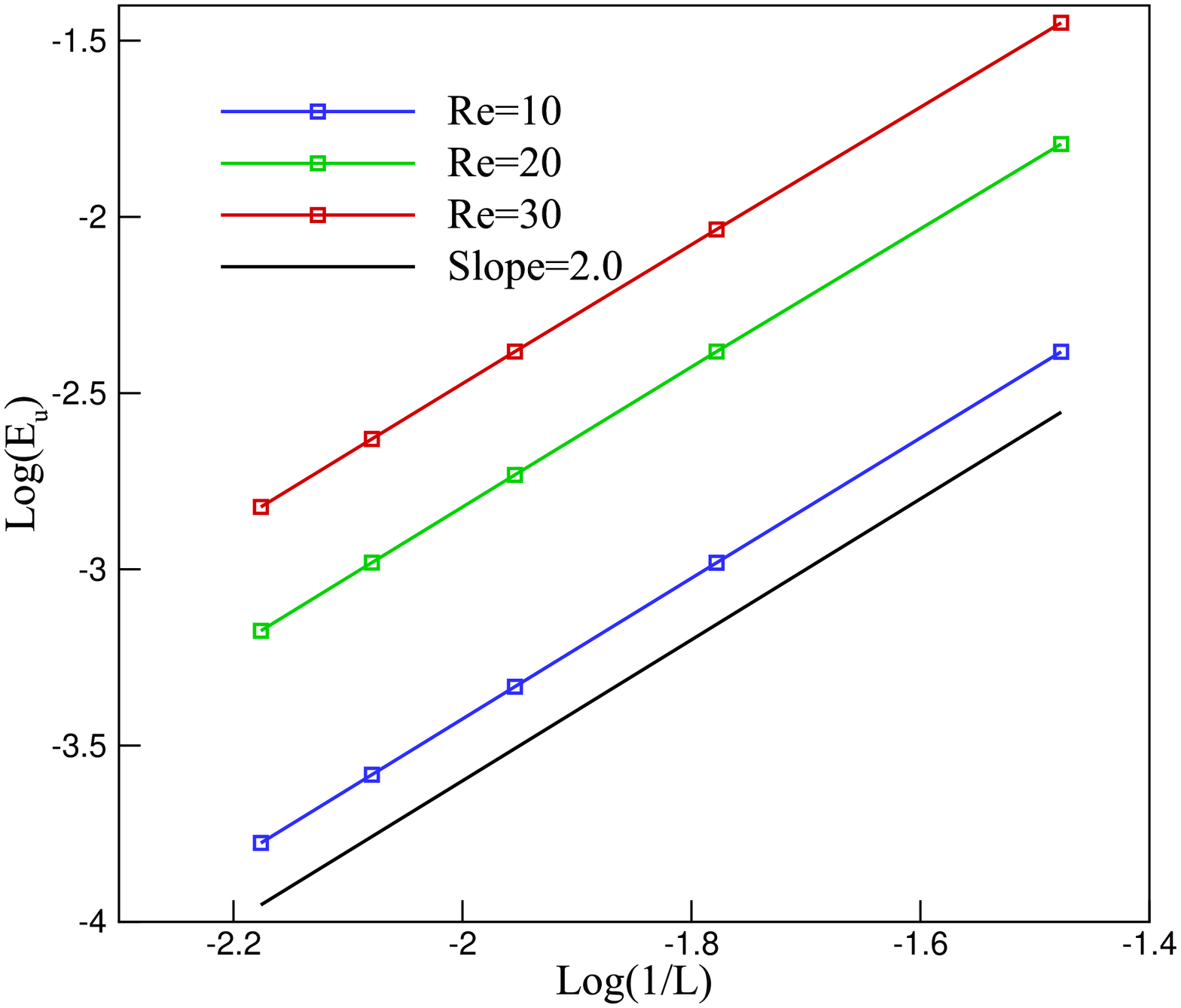,width=0.4\textwidth,clip=}}\vspace{0.0cm}\\
	%	(a) $ Ra = {10^3}$\hspace{0.3\textwidth} 
	%	(b) $ Ra = {10^3}$\\
	}
	\caption{Relative errors of temperature (left) and velocity (right) change with lattice spacings at $Pr  =0.71 $ and $ Re  = [10,20,30] $.}
	\label{FIG2b}
\end{figure*}
The relationships between grid size and relative errors of the present method are plotted in Fig. \ref{FIG2b}, and the slopes of each fitting lines are very close to 2.0. This demonstrates that the method proposed is of second-order convergence rate in space.
\subsection{Rayleigh-B\'{e}nard convection}\label{sec.3.3}
In this section, the Rayleigh-B\'{e}nard convective flow is conducted to check the ability of simulating incompressible thermal flows with an external force field. In the 2D Rayleigh-B\'{e}nard convective flow, the fluid is enclosed between two parallel stationary walls, with high temperature $ T_H $ at the bottom and low temperature $ T_L $ at the top, and experiences the gravity. The gravity field is incorporated  by Eq. (\ref{e3}). 

The flow is characterized by the length-width ratio of the flow domain $ L:H $, the Prandtl number $ Pr  = \nu /\alpha $ and Rayleigh number $ Ra = g\beta \Delta T{H^3}/(\nu \alpha ) $. In the simulations, we set $L \times H = 60 \times 30 $, $ Pr  = 0.71 $, $ T_H=1.05 $, $ T_L=0.95 $,
and $ T_0=({T_H}+{T_L})/2=1.0 $. The characteristic velocity of the convection is $ {u_c} = \sqrt {g\beta \Delta TH} $, by which the Mach number is defined, $ Ma = {u_c}/{c_s} $. For a given Rayleigh number, the viscosity is calculated according to $\nu  = Ma{c_s}H\sqrt {\Pr /Ra} $. To set up the simulation, a initial small disturbance is given to the density field along the horizontal center line,
\begin{equation}
\rho (x,H/2) = {\rho _0}\left[ {1.0 + 0.001\cos (2\pi x/L)} \right],
\end{equation}
where $ {\rho _0} = 1.0 $, while other points are initialized as $ {\rho _0} $.

\begin{figure*}[!ht]
	\center {
		{\epsfig{file=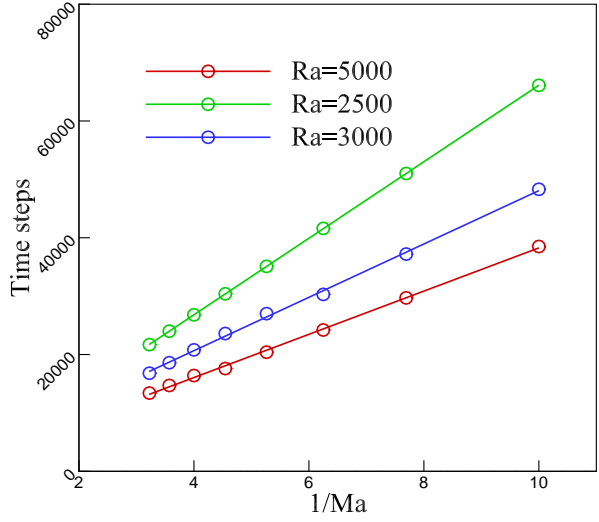,width=0.5\textwidth,clip=}}	
	}
	\caption{Time steps needed for convergence change with the reciprocal of $ Ma $.}
	\label{FIG3a}
\end{figure*}

The characteristic time of the system can be expressed by
$ {t_c} \sim H/{u_c} = H{c_s}/Ma $. It is known that the iterations needed for the convergence are proportional to $ {t_c} $. Firstly, we change $ 1/Ma $ from 0.1 to 0.31 with a 0.03 interval, and calculte the needed time steps until convergence. From Fig. \ref{FIG3a}, we certainly confirm the linear relation between the needed time steps and $1/Ma$. Throughout the variation range of $ Ma $, the relative changes of the Nusselt number
(defined in Eq. \ref{eq29}) are only $ {\rm{0}}{\rm{.11\% }}$, $ {\rm{0}}{\rm{.10\% }}$ and $ {\rm{0}}{\rm{.08\% }}$ for the Rayleigh numbers at 2500, 3000, and 5000, respectively. Based on this, a higher Mach number can be used in the present method to reduce the computational cost, but not affects the numerical accuracy. In the following simulations, the Mach number is set to $ Ma=0.3 $.

According to the linear stability theory, the driven force by the density variations induced by the temperature variations will be balanced by the viscous force when  Rayleigh number is lower than a critical value $ {{R}}{{{a}}_c} $, while if the Rayleigh number is increased above the threshold, the driving force will dominate and convection will be induced.
\begin{figure*}[!ht]
	\center {
		{\epsfig{file=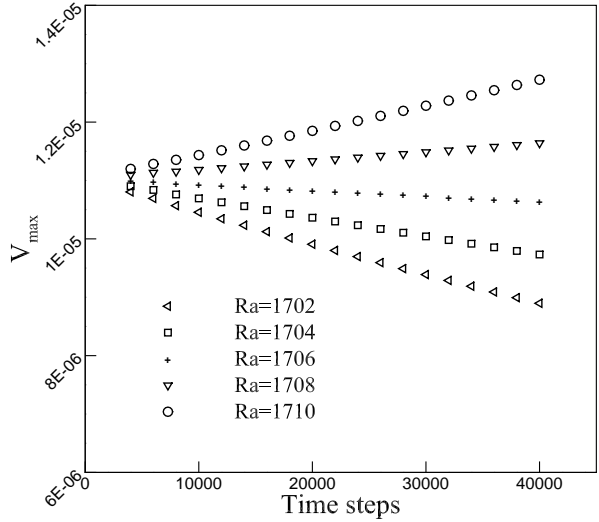,width=0.5\textwidth,clip=}}	
	}
	\caption{The evolution of $ {{\rm{V}}_{\max }} $ with time at $ {\rm{Ra}} = \left[ {1702,1704,1706,1708,1710} \right] $.}
	\label{FIG3b}
\end{figure*}
To determine the critical 
Rayleigh number, we measure the evolution of the maximun vertical velocity in the system $ {{{V}}_{\max }} $ for a series of Rayleigh numbers, $ {{Ra}} = \left[ {1702,1704,1706,1708,1710} \right] $. It can be seen in Fig \ref{FIG3b} that $ {{{V}}_{\max }} $ will keep increasing/decreasing approximately linearly in the early period, depending on $ {{R}}{{{a}}} $. The critical Rayleigh number is determined by solving zero value for the grow rate of $ {{{V}}_{\max }} $ with the least-squre method \cite{wang2013lattice}. Compared with the exact value in the linear stability theory, $ {{R}}{{{a}}_c}=1707.76 $, the value based on our method, 1706.82, is satisfying.

\begin{figure*}[!ht]
	\center {
		{\epsfig{file=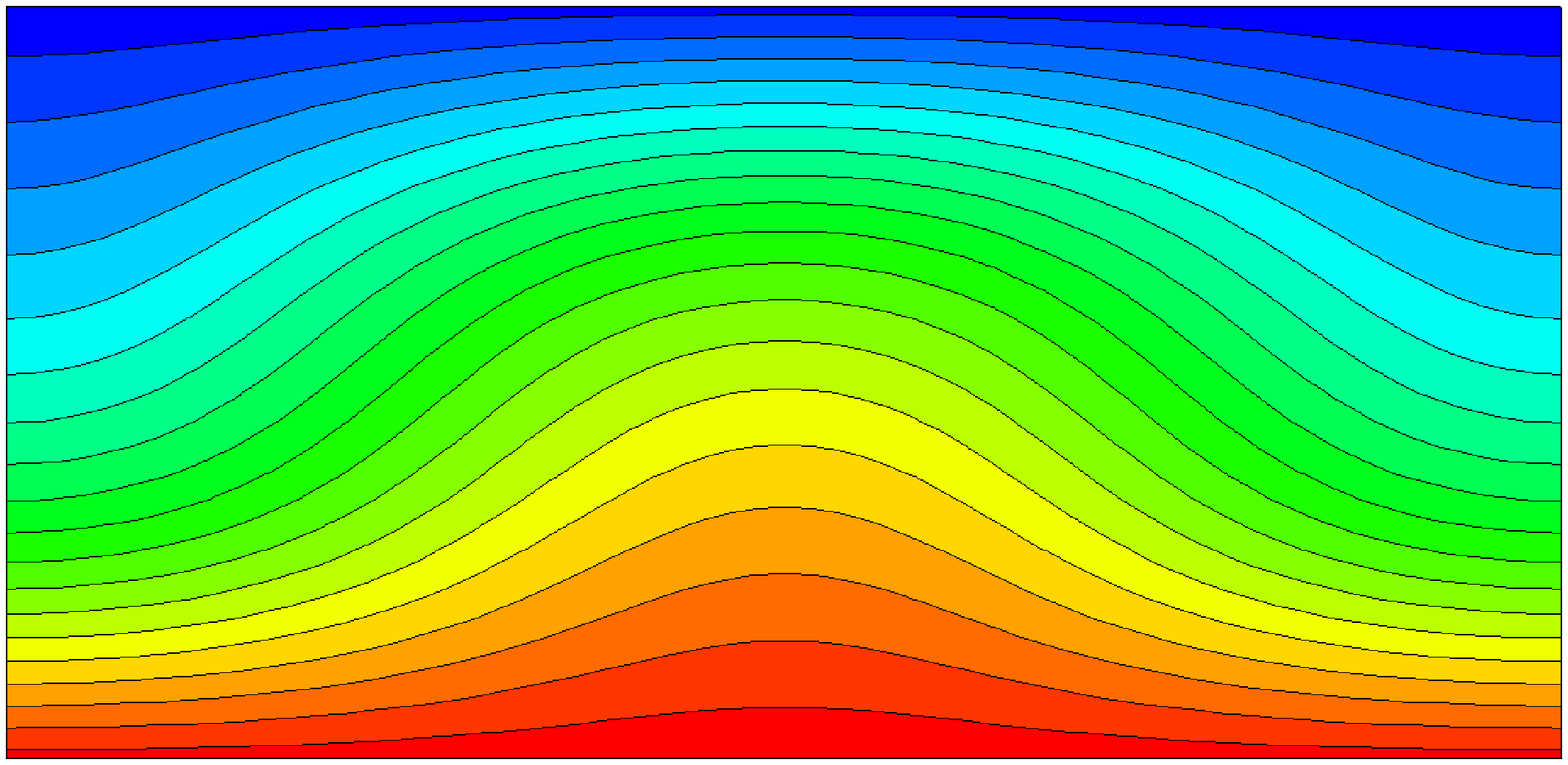,bbllx=15pt,bblly=15pt,bburx=700pt,bbury=350pt,width=0.6\textwidth,clip=}}\\ 
		{\epsfig{file=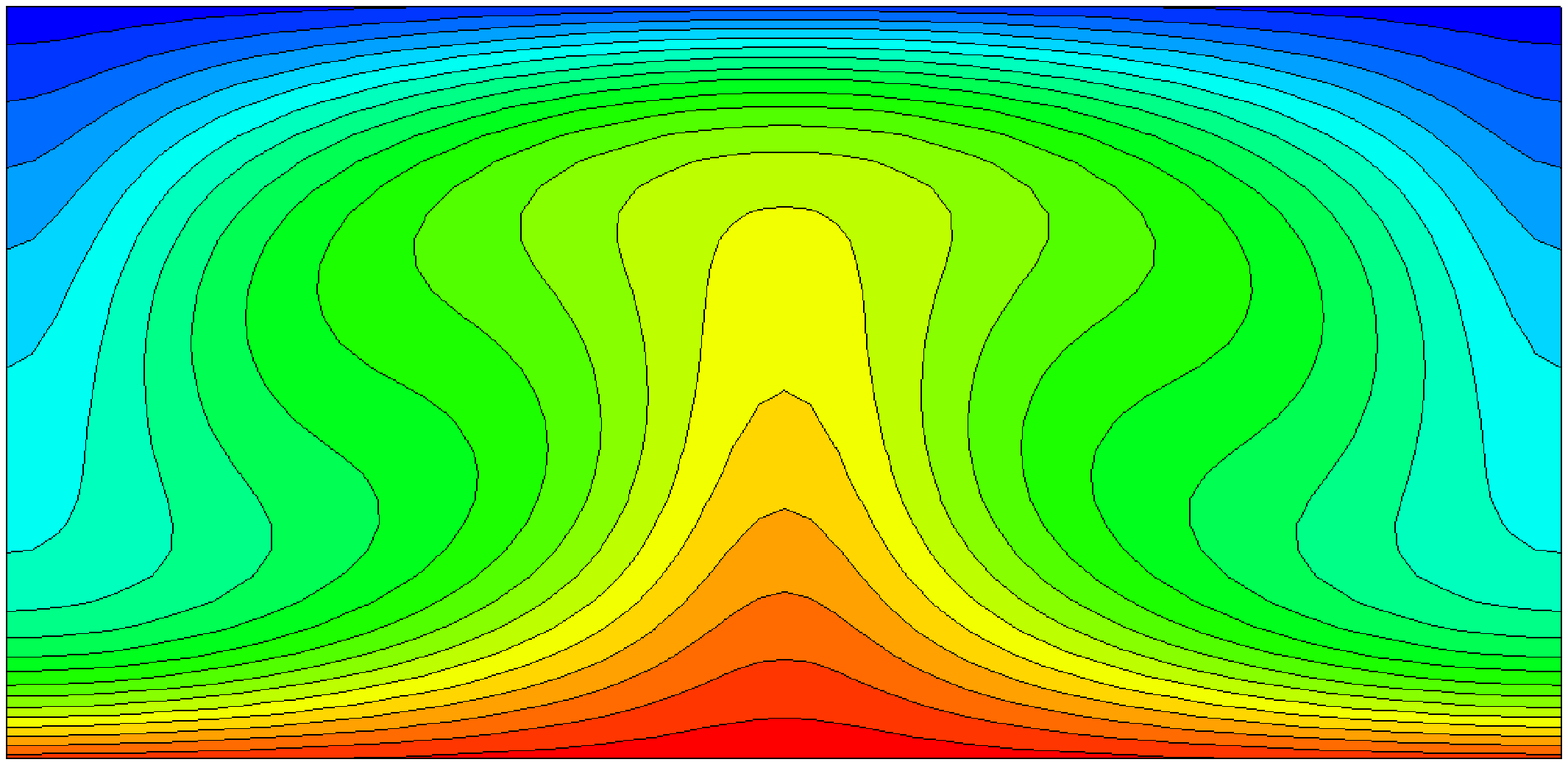,bbllx=15pt,bblly=15pt,bburx=700pt,bbury=350pt,width=0.6\textwidth,clip=}}\\
		{\epsfig{file=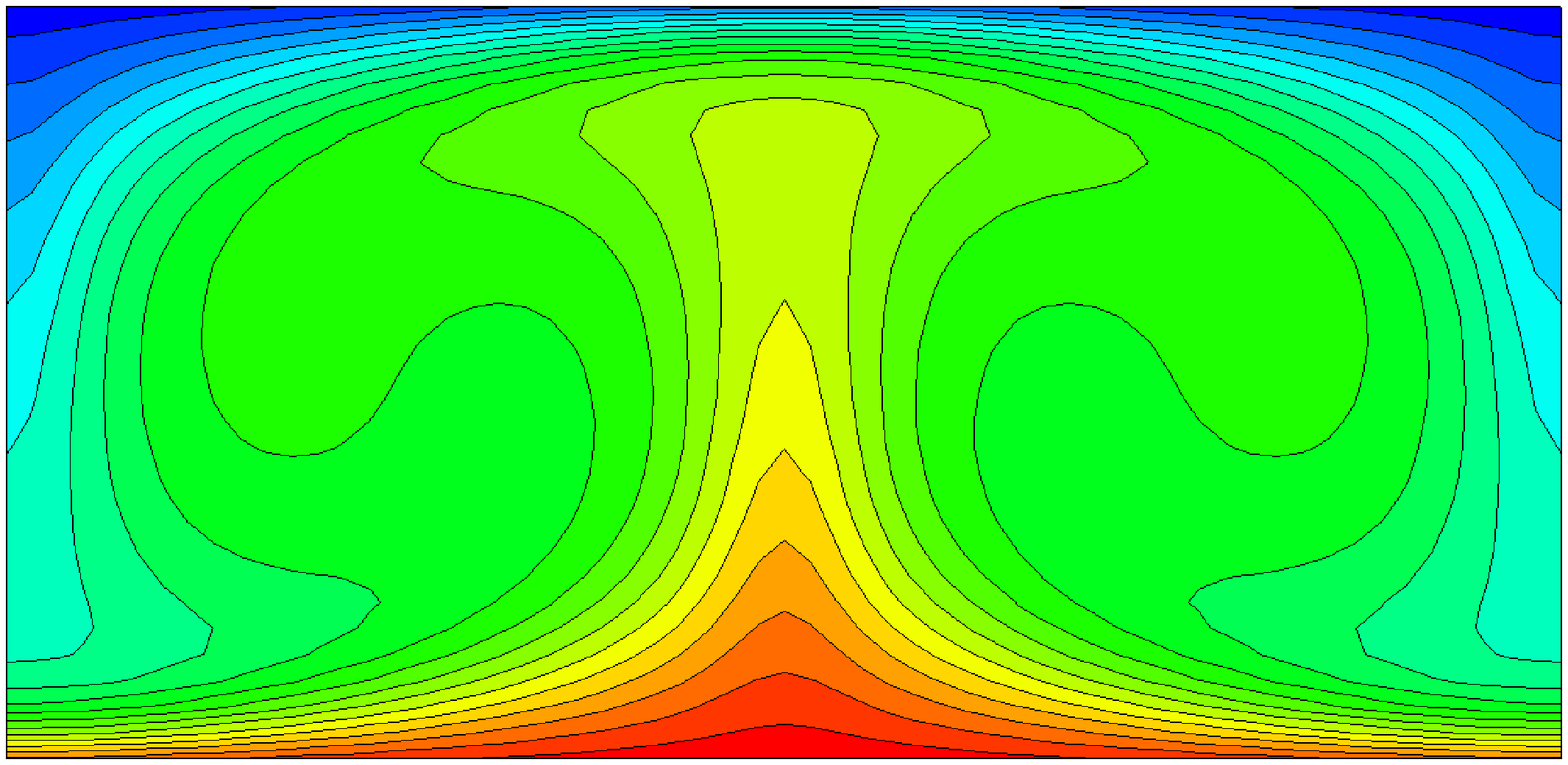,bbllx=15pt,bblly=15pt,bburx=700pt,bbury=350pt,width=0.6\textwidth,clip=}}\\  
	}
	\caption{Isotherms for Rayleigh-B\'{e}nard convective flows. From top to bottom,  $ Ra = {2000}$, 10000 and 50000.}
	\label{FIG04}
\end{figure*}
\begin{figure*}[!ht]
	\center {
		{\epsfig{file=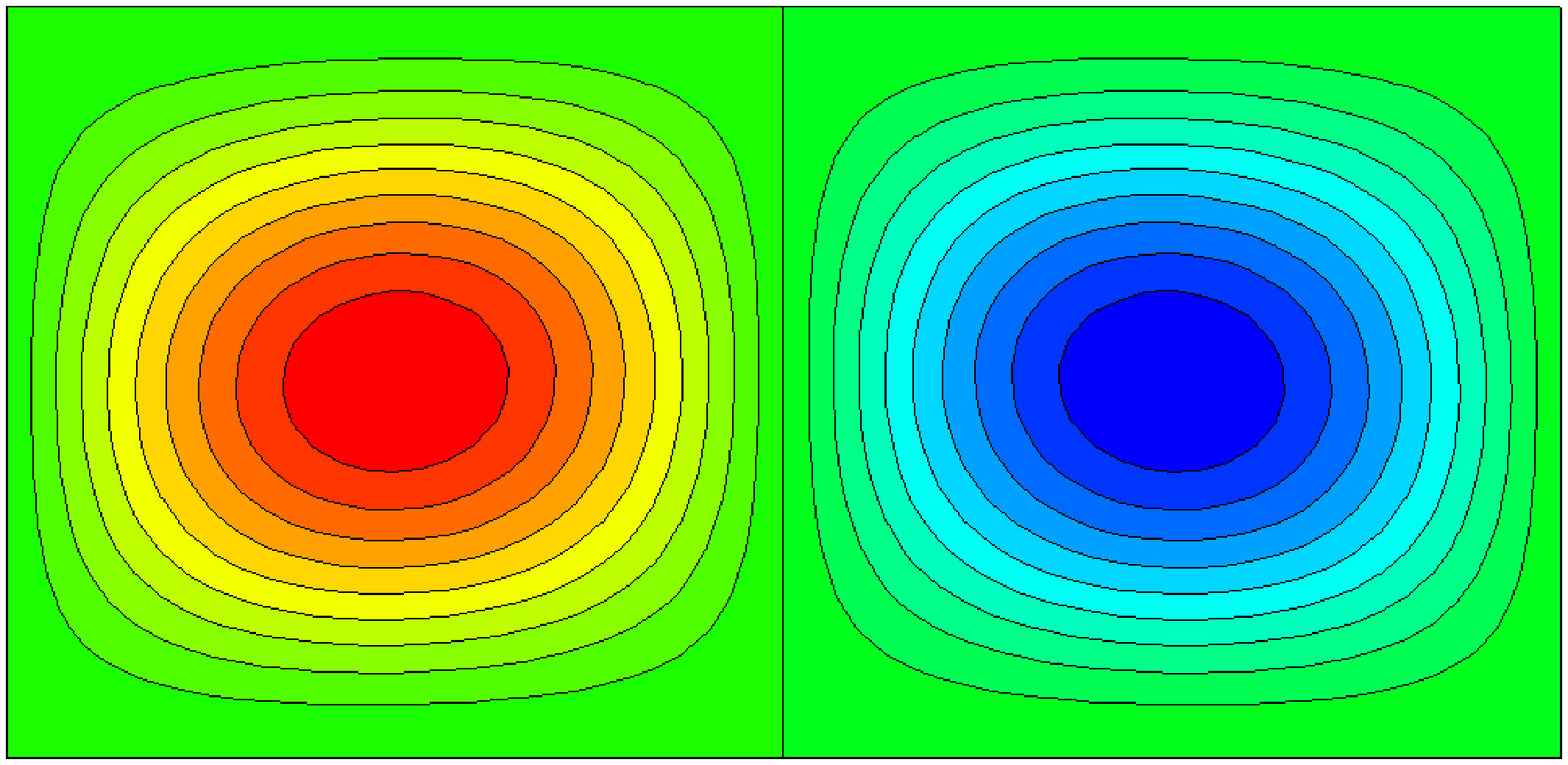,bbllx=15pt,bblly=15pt,bburx=700pt,bbury=350pt,width=0.6\textwidth,clip=}}\\ 
		{\epsfig{file=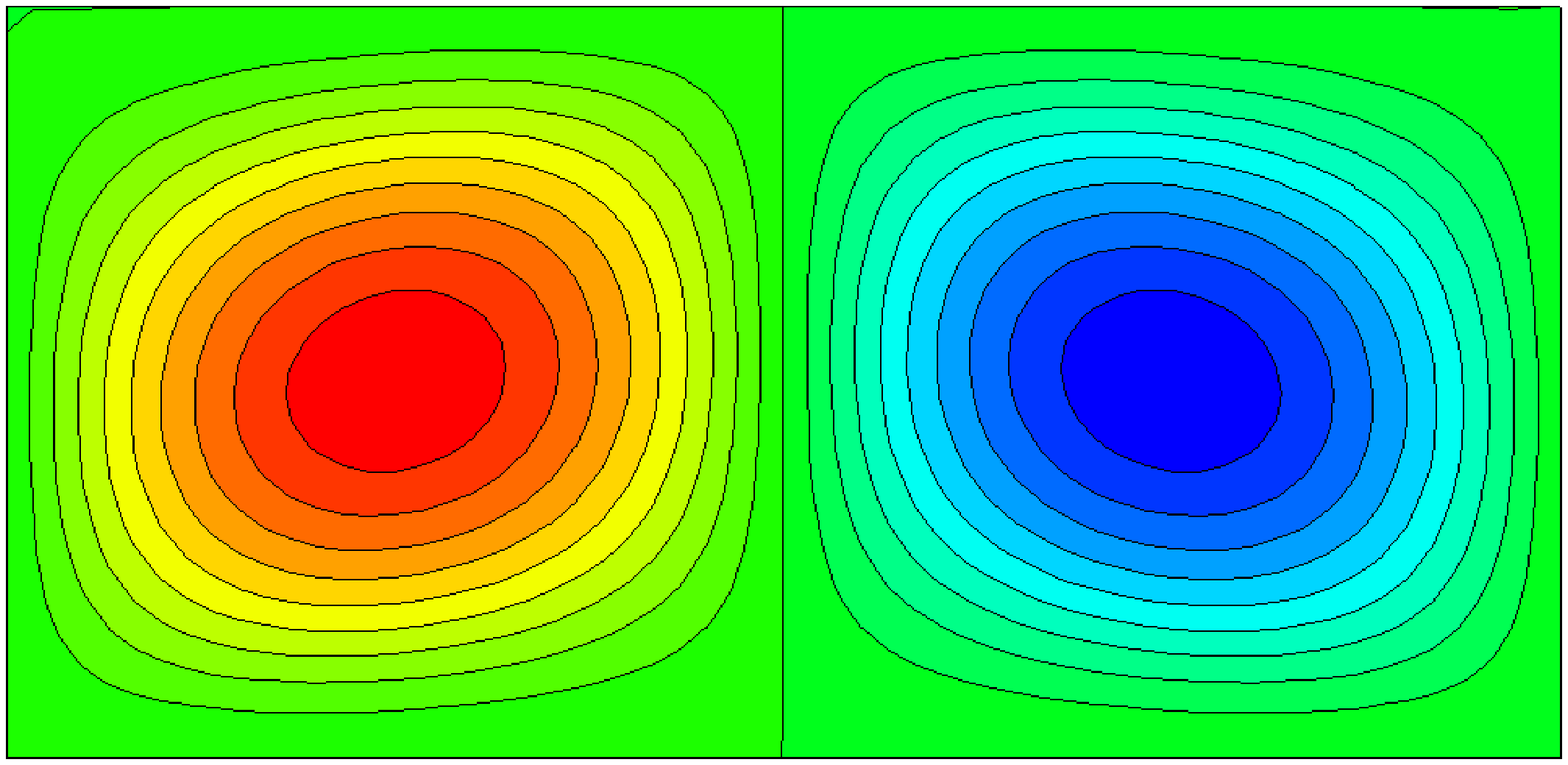,bbllx=15pt,bblly=15pt,bburx=700pt,bbury=350pt,width=0.6\textwidth,clip=}}\\
		{\epsfig{file=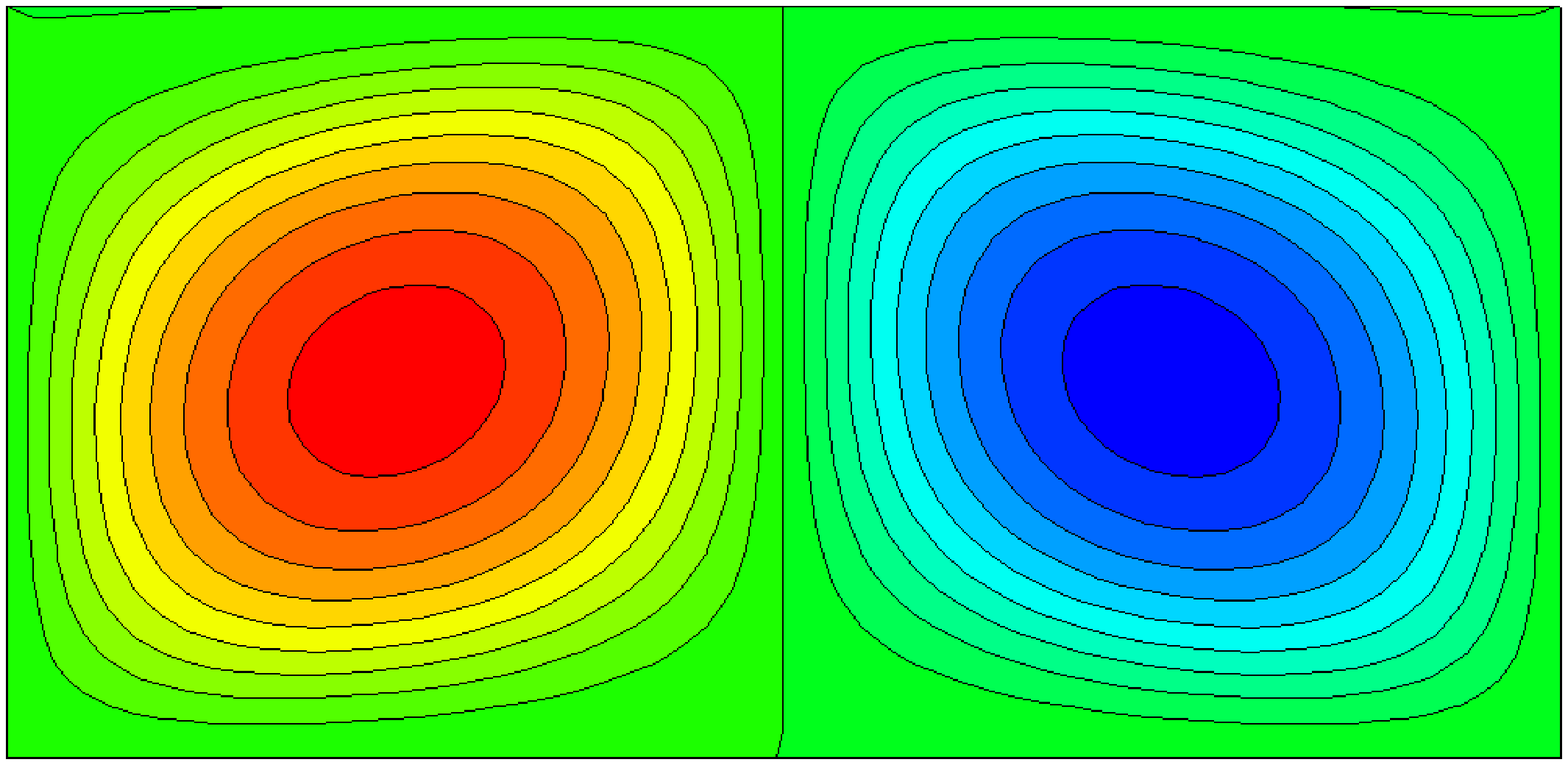,bbllx=15pt,bblly=15pt,bburx=700pt,bbury=350pt,width=0.6\textwidth,clip=}}\\  
	}
	\caption{Streamlines for Rayleigh-B\'{e}nard convective flows. From top to bottom,  $ Ra = {2000}$, 10000 and 50000.}
	\label{FIG05}
\end{figure*}
Flows at different Rayleigh numbers are then simulated. Figs. \ref{FIG04} display the normalized temperature $ \left( {T - {T_0}} \right)/\Delta T $ at $ Ra = {2000}$, 10000 and 50000. When the Rayleigh number increases, we can see two clear trends in the figures: the mixing of the hot and cold fluids is enhanced, and the temperature gradients near the bottom and top walls are increased, both of which mean the convective heat transfer is enhanced in the domain. In the meantime, as shown in Figs .\ref{FIG05}, the vortex is gradually distorted with the increase of the Rayleigh number, which also means the enhancement of convection.To quantify this, the Nusselt number in the system is calculated,
\begin{equation}\label{eq29}
Nu = 1 + \frac{{\left\langle {{u_y}T} \right\rangle H}}{{\alpha \Delta T}},
\end{equation}
where the square bracket represents the average over the whole system. Nusselt numbers obtained at various Rayleigh numbers are compared with the reference data in Table \ref{TAB1}. The simulation results are in good agreement with the analytical values in Ref .\cite{clever1974transition}. 
\begin{table*}
	\renewcommand\arraystretch{1.3}
	\caption{\label{TAB1} Comparison of Nusselt number between the present numerical results and the results in Ref. \cite{clever1974transition,prasianakis2007lattice}.
	}
\centering
		\begin{tabular}{cccccc}
			\toprule
			& &\multicolumn{2}{c}{Present method }  &\multicolumn{2}{c}{Results in Ref .\cite{prasianakis2007lattice}}  
			\\ 
			\cline{3-4} \cline{5-6}			
			 Cases
			& Analytic $N_u$ \cite{clever1974transition} & $N_u$ &Relative error&$N_u$ &Relative error
			\\ 
			\hline			
			2000&1.212&1.213&0.08&-&-\\	
			2500&1.475&1.477&0.13&1.474&0.07\\
			3000&1.663&1.667&0.24&-&-\\
			5000&2.116&2.121&0.24&2.104&0.57\\	
			10000&2.661&2.672&0.41&2.664&0.64\\	
			20000&3.258&3.271&0.4&-&-\\
			30000&3.662&3.668&0.16&3.605&1.56\\
			50000&4.245&4.229&0.38&4.133&2.64\\	
		\toprule		
		\end{tabular}
\end{table*}
On the whole, the relative errors for the present method are smaller than the method in \cite{prasianakis2007lattice}, while only 30 nodes are used in the present method rather than 50 nodes as in \cite{prasianakis2007lattice}.
\begin{figure*}[!ht]
	\center {
		{\epsfig{file=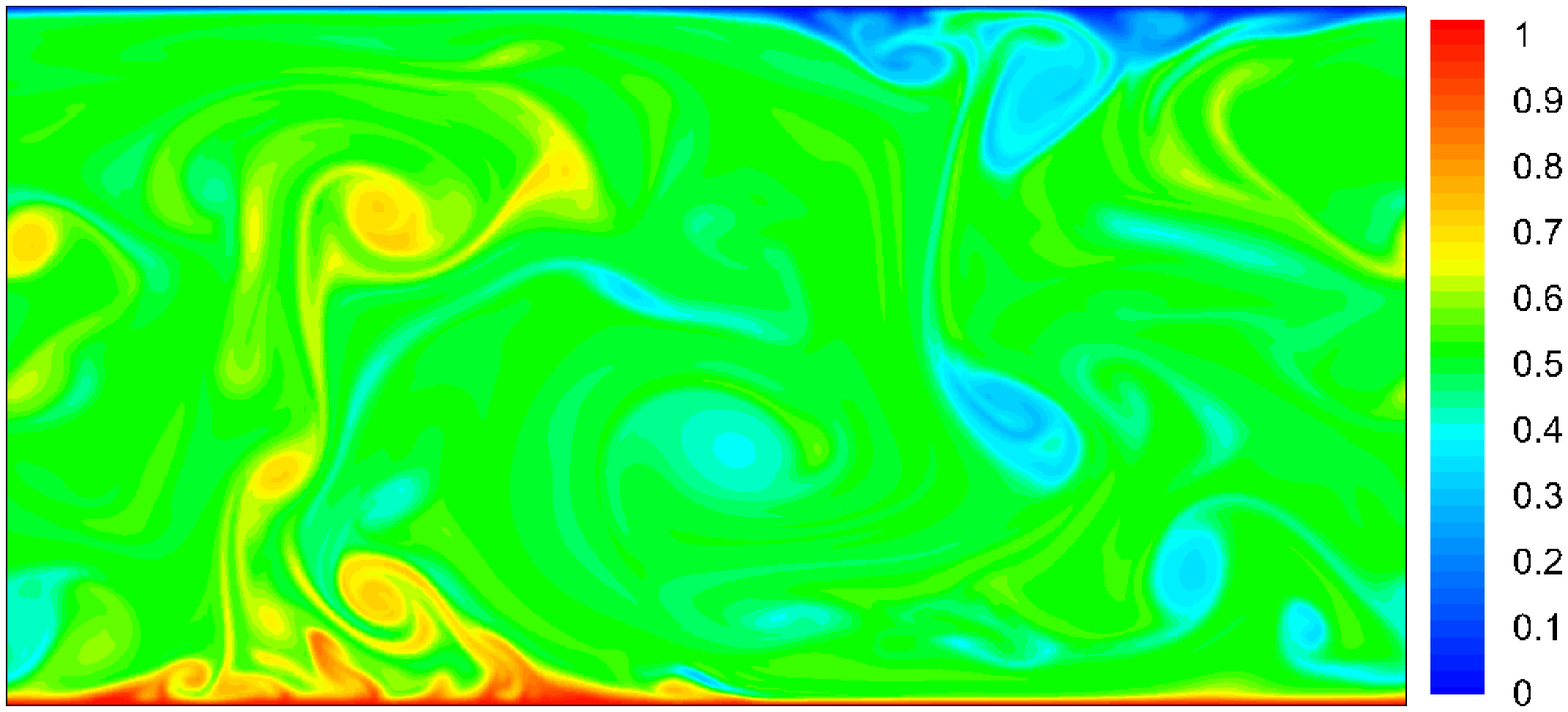,bbllx=15pt,bblly=15pt,bburx=650pt,bbury=320pt,width=0.7\textwidth,clip=}}	
	}
	\caption{Snapshot of the temperature field at ${Ra=10^9}$.}
	\label{FIG06}
\end{figure*}

In the end, it is interesting to find that the proposed method is stable for $Ra$ reaching up to ${10^9}$ with only 220 nodes in the vertical direction as shown in Fig .\ref{FIG06}, which confirms the commendable stability of the method. However, the study of high-Rayleigh-number thermal convection is beyond the scope of the current paper.
\section{Conclusions}\label{sec.4}
In this paper, we have developed a central-moment-based lattice Boltzmann (CLB) model for simulation of incompressible thermal flows. Combined with the D2Q9 CLB equation for the flow field, another D2Q5 CLB equation is designed to reproduce the temperature equation. Through the Chapman-Enskog analysis, the macroscopic govering equations for incompressible thermal flows can be recovered. By matching the higher order central moments of continuous Maxwell-Boltzmann distribution naturally, the proposed thermal model achieves better Galilean invariance compared with some existing thermal LBMs. Thus a higher Mach number can be adopted for convection flows in the present model, which reduces the computational cost significantly. Through the simulation of several canonical problems, the accuray, efficiency, stability and the second-order convergence rate in space for the present model are verified. The method developed retains the simplicty and numerical efficiency of the standard LBM. In principle, the method can be extended to three-dimensional readily. Besides, the model developed can also be applied to other convection-diffusion problems directly. 
\section *{Acknowledgments}

Support from the MOST National Key Research and Development Programme (Project No. 2016YFB0600805) and the Center for Combustion Energy at Tsinghua University is gratefully acknowledged. The simulations were partly performed on the Tsinghua High-Performance Parallel Computer supported by the Tsinghua National Laboratory for Information Science and Technology and partly on ARCHER funded under the EPSRC project ``UK Consortium on Mesoscale Engineering Sciences (UKCOMES)" (Grant No. EP/L00030X/1). 

\section*{Appendix A. Chapman-Enskog analysis of the CLBM for temperature field}
The explicit expressions for $  {\bf{M}}_\textbf{T}^{ - 1} $ and $ {\bf{N}}_\textbf{T}^{ - 1} $ are as follows.
\begin{equation}
{\bf{M}}_\textbf{T}^{ - 1}= \left[ 
\begin{array}{c c c c c}
1 &0 &0&-1&-1\\
0&1/2&0&1/2&0\\
0&0&1/2&0&1/2\\
0&-1/2&0&1/2&0\\
0&0&-1/2&0&1/2\\
\end{array} 
\right],
\end{equation}
\begin{equation}
{\bf{N}}_\textbf{T}^{ - 1} = \left[ 
\begin{array}{c c c c c}
1 &0 &0&0&0\\
{u_x}&1&0&0&0\\
{u_y}&0&1&0&0\\
u_x^2&2{u_x}&0&1&0\\
u_y^2&0&2{u_y}&0&1\\
\end{array} 
\right].
\end{equation} 
The collision step in Eq. (\ref{e19}) and streaming step in Eq. (\ref{e22}) and can be integratged into one step as,
\begin{equation}\label{e26}
{g_i}({\bf{x}} + {{\bf{e}}_i}\Delta t,t + \Delta t) - {g_i}({\bf{x}},t) =  - {\bf{M}}_\textbf{T}^{ - 1}{\bf{N}}_\textbf{T}^{ - 1}{{\bf{S}}_\textbf{T}}{{\bf{N}}_\textbf{T}}{{\bf{M}}_\textbf{T}}\left| {{g_i} - g_i^{eq}} \right\rangle. 
\end{equation} 
According to the Chapmann-Enskog expansion, the following multiscale expansions are usually introduced,
\begin{subequations}\label{e27}
	\begin{equation}
	{g_i}({\bf{x}} + {{\bf{e}}_i}\Delta t,t + \Delta t) = \sum\nolimits_{i = 0}^\infty  {\frac{{{\varepsilon ^n}}}{{n!}}({\partial _t} + } {{\bf{e}}_i} \cdot \nabla {)^n}{g_i}({\bf{x}},t), 
	\end{equation} 
	\begin{equation}
	{g_i} = g_i^{(0)} + \varepsilon g_i^{(1)} + {\varepsilon ^2}g_i^{(2)} + ...,~~
	{\partial _t} = \varepsilon {\partial _{t1}} + {\varepsilon ^2}{\partial _{t2}},~~
	\nabla  = \varepsilon {\nabla _i},	
	\end{equation} 
\end{subequations}
where $ \varepsilon $ is the expansion parameter. Using these expansions, the D2Q5 CLBM in Eq. (\ref{e26}) can be written in the consecutive orders of $ \varepsilon $,
\begin{subequations}\label{e28}
	\begin{equation}
	O({\varepsilon ^0}):g_i^{(0)} = g_i^{eq},
	\end{equation} 
	\begin{equation}
	O({\varepsilon ^1}):({\partial _{t1}} + {{\bf{e}}_i} \cdot {\nabla _1})g_i^{(0)} =  - \frac{1}{{\Delta t}}{\bf{M}}_\textbf{T}^{ - 1}{\bf{N}}_\textbf{T}^{ - 1}{{\bf{S}}_\textbf{T}}{{\bf{N}}_\textbf{T}}{{\bf{M}}_\textbf{T}}\left| {g_i^{(1)}} \right\rangle ,	
	\end{equation} 
	\begin{equation}
	\begin{aligned}
	O({\varepsilon ^2}):{\partial _{t2}}g_i^{(0)} + ({\partial _{t1}} + {{\bf{e}}_i} \cdot {\nabla _1})g_i^{(1)} + \frac{{\Delta t}}{2}{({\partial _{t1}} + {{\bf{e}}_i} \cdot {\nabla _1})^2}g_i^{(0)} =\\
	 - \frac{1}{{\Delta t}}{\bf{M}}_\textbf{T}^{ - 1}{\bf{N}}_\textbf{T}^{ - 1}{{\bf{S}}_\textbf{T}}{{\bf{N}}_\textbf{T}}{{\bf{M}}_\textbf{T}}\left| {g_i^{(2)}} \right\rangle.
	 \end{aligned}	
	\end{equation} 
\end{subequations}
If we multiply the matrix $ {{\bf{M}}_\textbf{T}} $ on both sides of Eqs. (\ref{e28}), the corresponding equations in the raw moment can be rewritten as,
\begin{subequations}\label{e29}
	\begin{equation}
	O({\varepsilon ^0}):\Gamma _i^{T,(0)} = \Gamma _i^{T,eq},
	\end{equation} 
	\begin{equation}\label{e29b}
	O({\varepsilon ^1}):({\bf{I}}{\partial _{t1}} + {{\bf{E}}_x}{\partial _{x1}} + {{\bf{E}}_y}{\partial _{y1}})\Gamma _i^{T,(0)} =  - \frac{1}{{\Delta t}}{\bf{N}}_\textbf{T}^{ - 1}{{\bf{S}}_\textbf{T}}{{\bf{N}}_\textbf{T}}\Gamma _i^{T,(1)},	
	\end{equation} 
	\begin{equation}
	\begin{aligned}
	O({\varepsilon ^2}):{\partial _{t2}}\Gamma _i^{T,(0)} + ({\bf{I}}{\partial _{t1}} + {{\bf{E}}_x}{\partial _{x1}} + {{\bf{E}}_y}{\partial _{y1}})({\bf{I}} - \frac{1}{2}{\bf{N}}_\textbf{T}^{ - 1}{{\bf{S}}_\textbf{T}}{{\bf{N}}_\textbf{T}})\Gamma _i^{T,(1)}= \\
 - \frac{1}{{\Delta t}}{\bf{N}}_\textbf{T}^{ - 1}{{\bf{S}}_\textbf{T}}{{\bf{N}}_\textbf{T}}\Gamma _i^{T,(2)},
	\end{aligned}
	\end{equation} 
\end{subequations}
where $ {{\bf{E}}_i} = {\bf{M}}_\textbf{T}[diag({e_{0i}},{e_{1,i}},...,{e_{4i}})]{{\bf{M}}_\textbf{T}^{ - 1}}(i = x,y) $ can be written explicitly as
\begin{subequations}
\begin{equation}
{{\bf{E}}_x}= \left[ 
\begin{array}{c c c c c}
0 &1 &0&0&0\\
0&0&0&1&0\\
0&0&0&0&0\\
0&1&0&0&0\\
0&0&0&0&0\\
\end{array} 
\right],
\end{equation}
\begin{equation}
{{\bf{E}}_y}= \left[ 
\begin{array}{c c c c c}
0 &0 &1&0&0\\
0&0&0&0&0\\
0&0&0&0&1\\
0&0&0&0&0\\
0&0&1&0&0\\
\end{array} 
\right].
\end{equation}
\end{subequations}
Writting out the equations for the conserved raw moment $\Gamma _0^T$, the following equations can be obtained,
\begin{subequations}\label{e30}
	\begin{equation}\label{e30a}
	O({\varepsilon ^0}):\Gamma _0^{T,(0)} = \Gamma _0^{T,eq},
	\end{equation} 
	\begin{equation}\label{e30b}
	O({\varepsilon ^1}):{\partial _{t1}}\Gamma _0^{T,(0)} + {\partial _{x1}}\Gamma _1^{T,(0)} + {\partial _{y1}}\Gamma _2^{T,(0)} =  - \frac{{{\lambda _0}}}{{\Delta t}}\Gamma _0^{T,(1)},	
	\end{equation} 
	\begin{equation}\label{e30c}
	\begin{aligned}
	O({\varepsilon ^2}):{\partial _{t2}}\Gamma _0^{T,(0)} + {\partial _{t1}}\left[ {(1 - \frac{{{\lambda _0}}}{2})\Gamma _0^{T,(1)}} \right] + {\partial _{x1}}\left[ {(1 - \frac{{{\lambda _1}}}{2})\Gamma _1^{T,(1)}} \right] +\\ {\partial _{y1}}\left[ {(1 - \frac{{{\lambda _1}}}{2})\Gamma _2^{T,(1)}} \right] =  - \frac{{{\lambda _0}}}{{\Delta t}}\Gamma _0^{T,(2)}.
	\end{aligned}
	\end{equation} 
\end{subequations}
According to Eq. (\ref{e30a}), we have $ \Gamma _0^{T,(n)} = 0 (n > 0)$. From Eq. (\ref{e29b}), we can get,
\begin{subequations}\label{e31}
	\begin{equation}
	\Gamma _1^{T,(1)} =  - \frac{{\Delta t}}{{{\lambda _1}}}\left[ {{\partial _{t1}}(T{u_x}) + {\partial _{x1}}(Tc_{T,s}^2 + Tu_x^2)} \right],
	\end{equation} 
	\begin{equation}
	\Gamma _2^{T,(1)} =  - \frac{{\Delta t}}{{{\lambda _1}}}\left[ {{\partial _{t1}}(T{u_y}) + {\partial _{y1}}(Tc_{T,s}^2 + Tu_y^2)} \right].	
	\end{equation} 
\end{subequations}
Substituting Eqs.(\ref{e31}) into Eq.(\ref{e30a}), we can get
\begin{equation}\label{e32}
O({\varepsilon ^2}):{\partial _{t2}}T - {\nabla _1}\left[ {\Delta t\left( {\begin{array}{*{20}{c}}
{1/{\lambda _1} - 0.5{\rm{,0}}}  \\
{{\rm{0,}}1/{\lambda _1} - 0.5}  \\
\end{array}} \right)\left( {\begin{array}{*{20}{c}}
{{\partial _{t1}}(T{u_x}) + {\partial _{x1}}(Tc_{T,s}^2 + Tu_x^2)}  \\
{{\partial _{t1}}(T{u_y}) + {\partial _{y1}}(Tc_{T,s}^2 + Tu_y^2)}  \\
\end{array}} \right)} \right] = 0.	
\end{equation}
Combining Eq.(\ref{e30b}) with Eq.(\ref{e32}) and Eq.(\ref{e1a}), we can obtain
\begin{equation}\label{e33}
{\partial _t}T + {\bf{u}} \cdot \nabla T = \nabla \cdot\left( {\alpha \nabla T} \right),
\end{equation}
where $ \alpha  = (1/{\lambda _1} - 0.5)c_{T,s}^2\Delta t $ is the thermal diffusion coefficient.

\section*{References}

\bibliography{mybibfile}
\end{document}